\documentclass[11pt,a4paper]{article}

\usepackage[T1]{fontenc}
\usepackage[utf8]{inputenc}
\usepackage{charter}
\usepackage[a4paper,margin=22mm]{geometry}
\usepackage{microtype}
\usepackage{graphicx}
\usepackage{float}
\usepackage{booktabs}
\usepackage{tabularx}
\usepackage{array}
\usepackage{amsmath}
\usepackage{enumitem}
\usepackage[font=small,labelfont=bf,labelsep=period]{caption}
\usepackage[authoryear,round]{natbib}
\usepackage{xurl}
\usepackage[hidelinks]{hyperref}
\usepackage{bookmark}
\usepackage{titlesec}

\graphicspath{{media/media/}}

\setlist[itemize]{leftmargin=1.4em,itemsep=0.35em,topsep=0.25em}

\bibpunct{(}{)}{;}{a}{,}{,}
\titleformat{\section}
  {\large\bfseries}
  {}
  {0pt}
  {}
\titlespacing*{\section}{0pt}{1.1em}{0.35em}
\titleformat{\subsection}
  {\normalsize\bfseries}
  {}
  {0pt}
  {}
\titlespacing*{\subsection}{0pt}{0.9em}{0.25em}

\newcolumntype{Y}{>{\raggedright\arraybackslash}X}

\begin{document}

\begin{center}
{\LARGE\bfseries SoilWaterNow: Soil water nowcasting for mapping plant available water (PAW) across paddocks for improved on-farm decision-making\par}
\vspace{0.7em}
{\large\itshape Yi Yu, Mikaela J. Tilse, Patrick Filippi and Thomas F. A. Bishop\par}
\vspace{0.25em}
{Precision Agriculture, Hydrology \& Geoinformation Science Laboratory, The University of Sydney\par}
\end{center}

\section{Keywords}

root-zone soil moisture; crop evapotranspiration; precision agriculture; yield potential

\section{GRDC code}

UOS2002-001RTX

\setlength{\fboxsep}{6pt}
\noindent\fbox{%
\begin{minipage}{\dimexpr\linewidth-2\fboxsep-2\fboxrule\relax}
\textbf{Take home message}
\begin{itemize}
\item We present the Sydney soil water-energy balance (SWEB) model, which provides a scalable, physically consistent modelling framework for translating satellite and climate data into actionable soil water information at 30 m resolution.
\item SWEB was validated nationally and demonstrated robust performance across Australia's diverse grain-growing environments.
\item Mid-season plant-available water nowcasts could be used in future to estimate water-limited yield potential, which could inform mid-season management decisions, such as nitrogen fertiliser top-up recommendations.
\end{itemize}
\end{minipage}}

\section{Background and rationale}

Water availability in the crop root zone is one of the primary determinants of crop growth, yield formation, and water-use efficiency across Australian grain-growing regions. While rainfall provides the primary input of water, crop performance is governed by the balance between root-zone soil moisture (RZSM), which represents water stored within the depth accessible to roots; and crop evapotranspiration (ET), which represents the combined water loss through plant transpiration and soil evaporation. Understanding both variables, and their interaction, is essential for effective monitoring of crop water status for better decision making.

In practice, substantial within-field variability in RZSM and ET is commonly observed due to spatial differences in soil texture, soil depth, topography, and historical management. These variations can lead to markedly different crop water stress responses within the same paddock, even under uniform rainfall conditions. However, many existing monitoring approaches fail to capture this heterogeneity at scales relevant to farm management. Point-based SM measurements provide direct information on RZSM but are spatially sparse and difficult to upscale, while regional satellite or model-based products often operate at resolutions too coarse to resolve paddock-scale variability.

Remote sensing has improved the ability to monitor actual ET at high spatial resolution, providing valuable insight into crop water use and stress conditions \citep{anderson_use_2012,senay_operational_2013}. ET integrates atmospheric demand, crop condition, and soil water availability, making it a powerful indicator of agricultural water stress. However, ET alone does not directly quantify how much water remains available to crops in the soil profile. Without an explicit representation of RZSM, it can be difficult to distinguish between short-term atmospheric effects and longer-term soil water limitations, particularly during critical growth stages. RZSM, by contrast, provides a direct measure of the water available for crop uptake and is strongly linked to yield variability and drought impacts \citep{mcnally_land_2017,seneviratne_investigating_2010}. Yet RZSM cannot be directly observed from satellites at the spatial and temporal resolutions required for agricultural decision-making. As a result, there is a clear need for modelling approaches that combine satellite-derived ET information with physically based representations of soil water dynamics, enabling consistent estimation of both ET and RZSM across scales.

To address this gap, the Sydney Soil Water-Energy Balance (SWEB) model was developed by explicitly linking surface energy processes (ET) with subsurface soil water balance processes (RZSM). By using satellite-informed estimates of actual ET as a key driver of soil water depletion, SWEB ensures that crop water use reflects real crop and atmospheric conditions. At the same time, the soil water balance component tracks infiltration, storage, redistribution, and drainage within the root zone, allowing estimation of how long available soil water can sustain crop demand.

For Australian agriculture and GRDC priorities, a scalable framework such as SWEB supports improved monitoring of within-field and regional water dynamics, enhanced understanding of spatial drivers of crop stress, and more informed management of water-limited systems. By jointly considering ET and RZSM, SWEB provides a physically consistent and operationally feasible approach for translating satellite and climate data into information that is directly relevant to crop water management, drought assessment, and productivity outcomes.

\begin{figure}[htbp]
\centering
\includegraphics[width=0.96\linewidth]{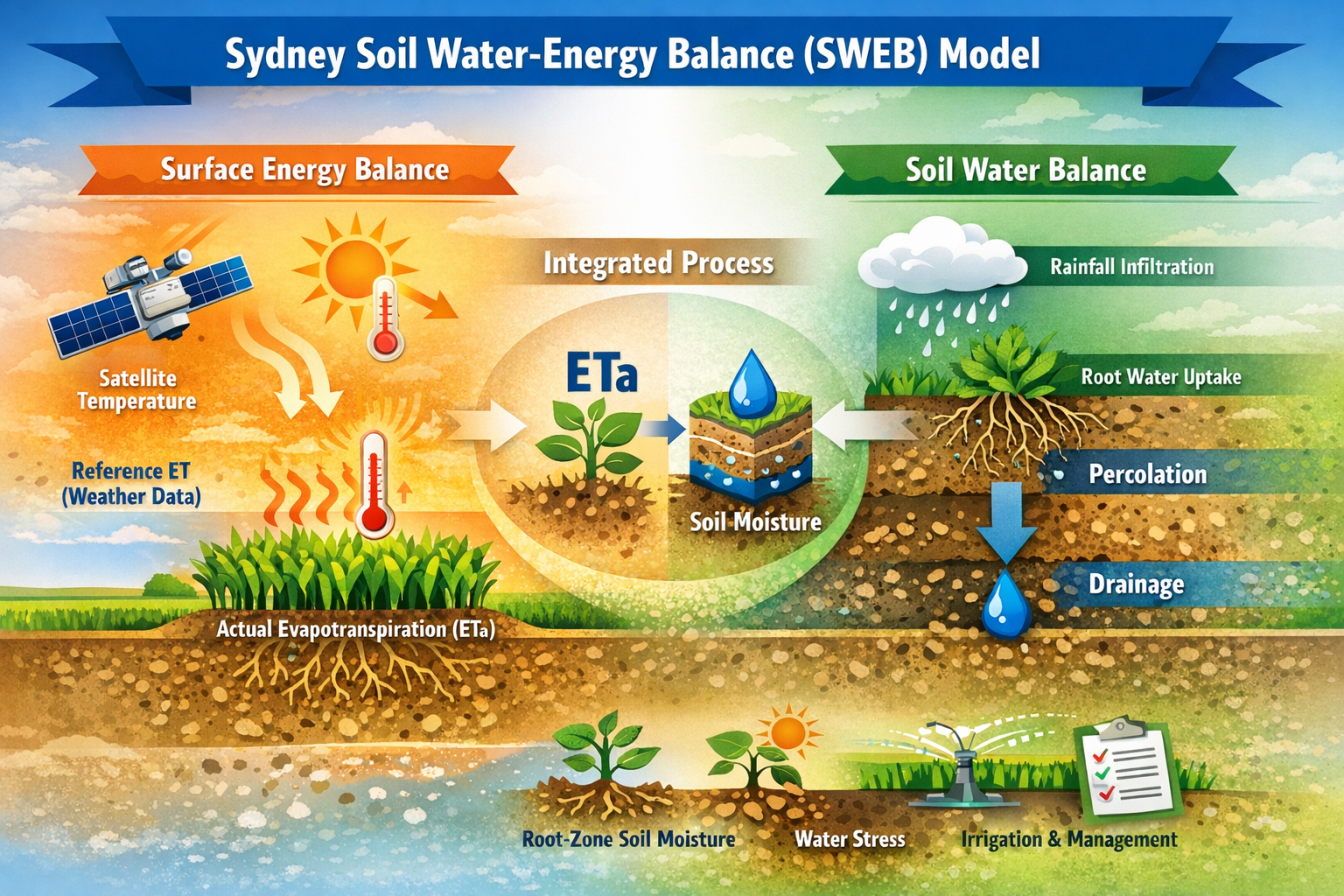}
\caption{Illustration of the SWEB components. Generated by ChatGPT and verified by the authors.}
\label{fig:sweb-components}
\end{figure}

\section{The Sydney soil water-energy balance (SWEB) model}

SWEB is a process-based modelling framework developed to estimate daily within-field crop ET and RZSM by integrating satellite observations, weather data, and soil physical properties. The model is designed to support agricultural water management by providing spatially explicit information on soil water availability and crop water stress across farming systems. SWEB consists of two tightly coupled components (Figure~\ref{fig:sweb-components}): (i) a surface energy balance (SEB) component that estimates actual ET, and (ii) a soil water balance (SWB) component that simulates water storage and movement within the soil profile.

\subsection{Surface energy balance (SEB) component: estimation of crop water use}

The energy component of SWEB estimates actual ET, which represents the total water loss from the land surface through plant transpiration and soil evaporation. This component is driven by satellite-derived land surface temperature (LST) from the Landsat 8/9 missions \citep{irons_next_2012,roy_next_2026}, which provide thermal observations at a spatial resolution (100 m, resampled to 30 m) suitable for resolving within-field variability in Australian grain systems. Landsat LST is combined with gridded meteorological inputs from the SILO database \citep{jeffrey_using_2001}, which supplies spatially consistent climate variables, including radiation, air temperature, vapour pressure, and FAO56 reference evapotranspiration ($\mathrm{ET}_{o}$). Together, these datasets enable physically consistent estimation of actual ET ($\mathrm{ET}_{a}$) that reflects both crop condition and atmospheric demand at paddock-relevant scales.

In principle, the model determines how much of the available atmospheric demand for water is actually met by the crop--soil system. When crops are well-watered, canopy temperatures tend to be lower and ET approaches the reference rate; under water stress, canopy temperatures increase and ET is reduced. The crop ET is estimated by scaling a reference ET, following the operational simplified surface energy balance (SSEBop) model \citep{senay_mapping_2022,senay_operational_2013}:

\begin{equation*}
\mathrm{ET}_{a} = K_{c} \times \mathrm{ET}_{o}
\tag{1}
\end{equation*}

\begin{equation*}
K_{c} = K_{cb} + K_{e}
\tag{2}
\end{equation*}

where $\mathrm{ET}_{o}$ is the reference FAO56 ET obtained from the SILO weather grids \citep{jeffrey_using_2001}; $K_{c}$ is the crop factor as described in \citet{allen_crop_1998}, which can be divided into two components---$K_{cb}$ (basal crop coefficient representing primarily plant transpiration) and $K_{e}$ (evaporation coefficient that represents the contribution of evaporation from soil to total ET). The resulting $\mathrm{ET}_{a}$ can be estimated at a resolution consistent with that of the input Landsat data, i.e., 30 m spatial resolution. The $\mathrm{ET}_{a}$ estimates represent actual water use, rather than potential demand, and therefore account for limitations imposed by SM availability, crop condition, and management (e.g., irrigation).

\subsection{Soil water balance (SWB) component: root-zone water dynamics}

The soil water balance component of SWEB tracks how water enters, moves through, and is stored within the soil profile. The soil is represented as a series of vertical layers extending from the surface to the maximum rooting depth.

In a simplified version, soil water storage on a daily time step is updated based on:

\begin{equation*}
\mathrm{RZSM}_{t+1} = \mathrm{RZSM}_{t} + I_{t} - \mathrm{ET}_{a,t} \cdot f_{evap} - D_{t}
\tag{3}
\end{equation*}

\begin{equation*}
I_{t} = P_{t} \cdot C_{infil}
\tag{4}
\end{equation*}

\begin{equation*}
f_{evap} = \frac{K_{e}}{K_{c}}
\tag{5}
\end{equation*}

where $\mathrm{RZSM}_{t}$ is RZSM at day $t$ (mm) and $\mathrm{RZSM}_{t+1}$ is the RZSM on the next day; $I_{t}$ is the rainfall infiltration into the soil (mm/day), which can be obtained from rainfall data $P_{t}$ together with a infiltration coefficient $C_{infil}$; the SEB component $\mathrm{ET}_{a,t}$ (mm/day) needs to be scaled by a soil evaporation fraction $f_{evap}$, which is the ratio between $K_{e}$ and $K_{c}$; $D_{t}$ is drainage/percolation out of the root zone (mm/day).

The model uses established soil physical relationships to represent how water moves between soil layers, accounting for soil texture, porosity, and hydraulic conductivity. SWEB is designed to be flexible with respect to soil inputs and can ingest any digital soil map (DSM) that provides basic textural information (e.g., sand and clay fractions) and soil depth. Soil hydraulic parameters are derived using pedotransfer functions (PTFs) \citep{saxton_soil_2006}, enabling consistent estimation of soil water retention and hydraulic conductivity across spatial scales. Upper limits on soil water storage ensure that soils cannot hold more water than their physical capacity, while drainage removes excess water beyond the root zone.

In this study, the Soil and Landscape Grid of Australia (SLGA; \citealp{malone_updating_2021,malone_updating_2021b}) is used as the baseline national soil dataset, allowing for the model to be applied anywhere. Where available, it is possible to use on-farm soil data such as digital soil maps (DSMs) generated from machine learning methods as developed in the USYD-led GRDC-funded project \emph{Next Generation Machine Learning models for 3D soil-mapping applications} (UOS2206-009RTX) providing spatially continuous coverage for regional applications. On-farm soil data provides more realistic representation of soil hydraulic properties at the on-farm scale.

\subsection{Integration of energy and water processes}

A key strength of SWEB is the explicit linkage between SEB (surface process) and SWB (subsurface process) dynamics. $\mathrm{ET}_{a}$ estimated from the SEB component is used as a direct sink of water in SWB. In this way, crop water use is constrained both by atmospheric conditions and by the amount of water actually available in the soil. This integration allows SWEB to capture important agricultural processes, such as monitoring crop water use during dry periods and waterlogging, and differences in water availability across soil types and landscapes.

\section{Calibration and operation of SWEB}

\subsection{Calibration settings}

SWEB needs to be spatially calibrated to ensure optimal within-field simulation performance. The calibration of SWEB employs a differential evolution optimisation algorithm to estimate three key parameters: $f_{evap}$, $C_{infil}$ and $k_{diff}$. The objective function minimises the root mean square error (RMSE) between observed and simulated SSM at 50 mm depth over a specified calibration period. Parameter bounds are constrained as follows:

\begin{itemize}
\item $f_{evap} \in [0,\,1]$ representing the fraction of total $\mathrm{ET}_{a}$ allocated to surface evaporation;
\item $C_{infil} \in [0,\,1]$ being a coefficient for precipitation infiltration into surface layer;
\item $k_{diff} \in [0,10^{6}]$ being a diffusivity scaling length factor in the Brooks-Corey formulation.
\end{itemize}

The optimisation process uses differential evolution with a population size of 10, maximum 30 iterations, and convergence tolerance of $10^{-3}$, requiring a minimum of 30 valid observation points for reliable parameter estimation. Currently SWEB is calibrated against surface SM retrievals from the Soil Moisture Active Passive (SMAP) mission \citep{entekhabi_soil_2010}, which is adjustable and can be calibrated to on-farm soil water sources such as SM probes.

\subsection{Operational settings through National Computational Infrastructure (NCI)}

SWEB has been deployed through NCI's Gadi supercomputer infrastructure to enable continental-scale RZSM modelling for Australia's key growing regions (Figure~\ref{fig:sm-network}). SWEB's Python-based \emph{ProcessPoolExecutor} architecture facilitates parallel execution across the national-scale SM monitoring sites, with each site calibration requiring independent optimisation runs that can be distributed across NCI's computing nodes. The outputs of SWEB could be distributed (subject to computational cost) as an open collection through the NCI Thematic Real-time Environmental Distributed Data Services (THREDDS) Data Service, providing supports for national agricultural and hydrological decision-makings.

\section{Validation of SWEB outputs}

\begin{figure}[H]
\centering
\includegraphics[width=0.8\linewidth]{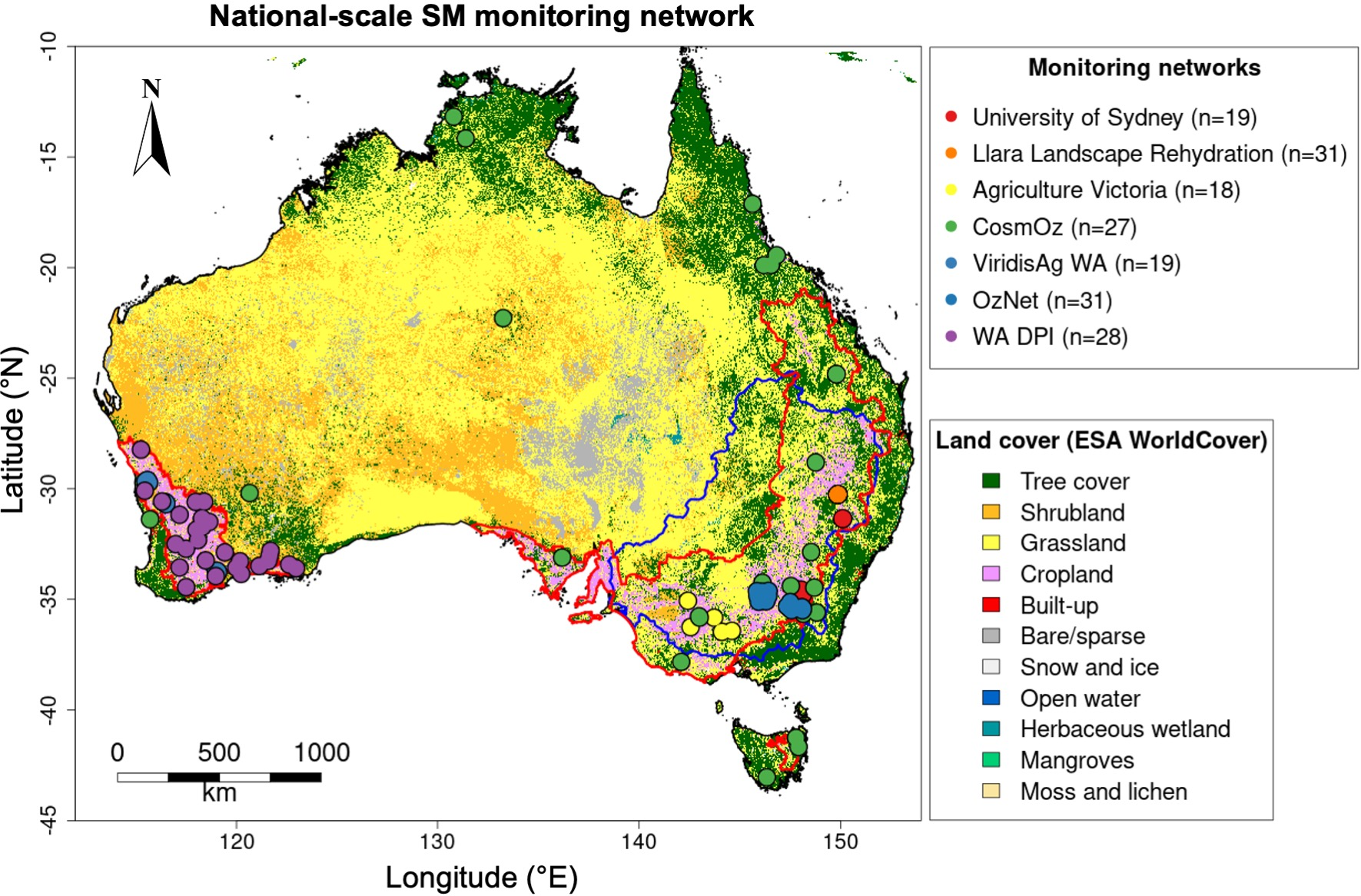}
\caption{The spatial distribution of a collated national-scale SM monitoring network for Australia. The dot points in various colours represent the sites from different SM monitoring networks. The land cover map is obtained from the ESA WorldCover 10 m v200 \citep{zanaga_esa_2022}. The red polygon delineates the grain growing regions. The blue polygon delineates the Murray Darling Basin.}
\label{fig:sm-network}
\end{figure}

SWEB is validated against a combination of soil moisture (SM) monitoring networks spanning key Australian grain-growing regions and climate zones (Table~\ref{tab:network-extent} and Figure~\ref{fig:sm-network}). These include:

\begin{itemize}
\item USYD Muttama network and the Murrumbidgee Soil Moisture Monitoring Network (MSMMN) in NSW, which provide profile-scale measurements of root-zone soil moisture across multiple depth intervals;
\item L'lara Landscape Rehydration network in northern NSW, representing field-scale variability within experimental farming systems;
\item VicAg network in Victoria and the WA DPI network in Western Australia, which offer regionally distributed profile soil moisture observations linked with local meteorological measurements;
\item CosmOz cosmic-ray network, which supplies area-integrated near-surface soil moisture estimates representative of paddock-scale conditions; and
\item ViridisAg WA network, providing on-farm soil moisture monitoring across multiple paddocks.
\end{itemize}

\begin{table}[htbp]
\centering
\caption{Spatial extent of each soil moisture monitoring network.}
\label{tab:network-extent}
\small
\begin{tabularx}{\linewidth}{Yrrrr}
\toprule
\textbf{Network} & \textbf{Min. Latitude} & \textbf{Min. Longitude} & \textbf{Max. Latitude} & \textbf{Max. Longitude} \\
\midrule
USYD & -34.84 & 147.96 & -30.255 & 150.124 \\
MSMMN & -35.497 & 145.849 & -34.621 & 148.132 \\
Llara Landscape Rehydration & -30.265 & 149.842 & -30.253 & 149.878 \\
VicAg & -37.775 & 141.614 & -34.454 & 145.748 \\
WA DPI & -34.451 & 115.26 & -28.231 & 122.958 \\
CosmOz & -43.042 & 115.71 & -13.179 & 149.801 \\
ViridisAg WA & -33.809 & 115.473 & -29.605 & 119.082 \\
\bottomrule
\end{tabularx}
\end{table}

While all sites within each network are documented, only stations meeting predefined data completeness and quality criteria---assessed in consultation with data custodians---are used for model evaluation. Together, these networks provide independent observations across a range of soil types, depths, spatial supports, and management contexts, enabling robust evaluation of SWEB-derived RZSM at a national scale.

\begin{figure}[htbp]
\centering
\includegraphics[width=\linewidth]{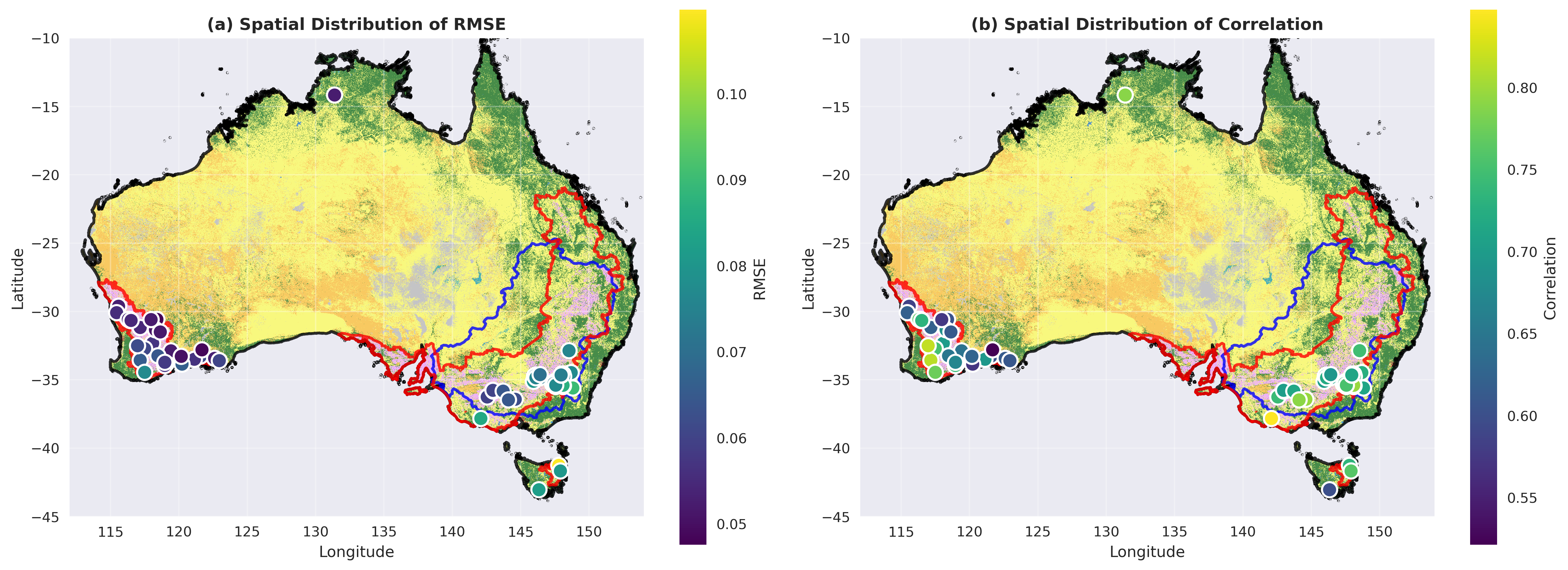}
\caption{National-scale spatial distribution of model performance metrics showing (a) RMSE and (b) correlation coefficient, with clear longitudinal gradients reflecting major climate zone transitions across Australia.}
\label{fig:metrics-map}
\end{figure}

\begin{figure}[H]
\centering
\includegraphics[width=\linewidth]{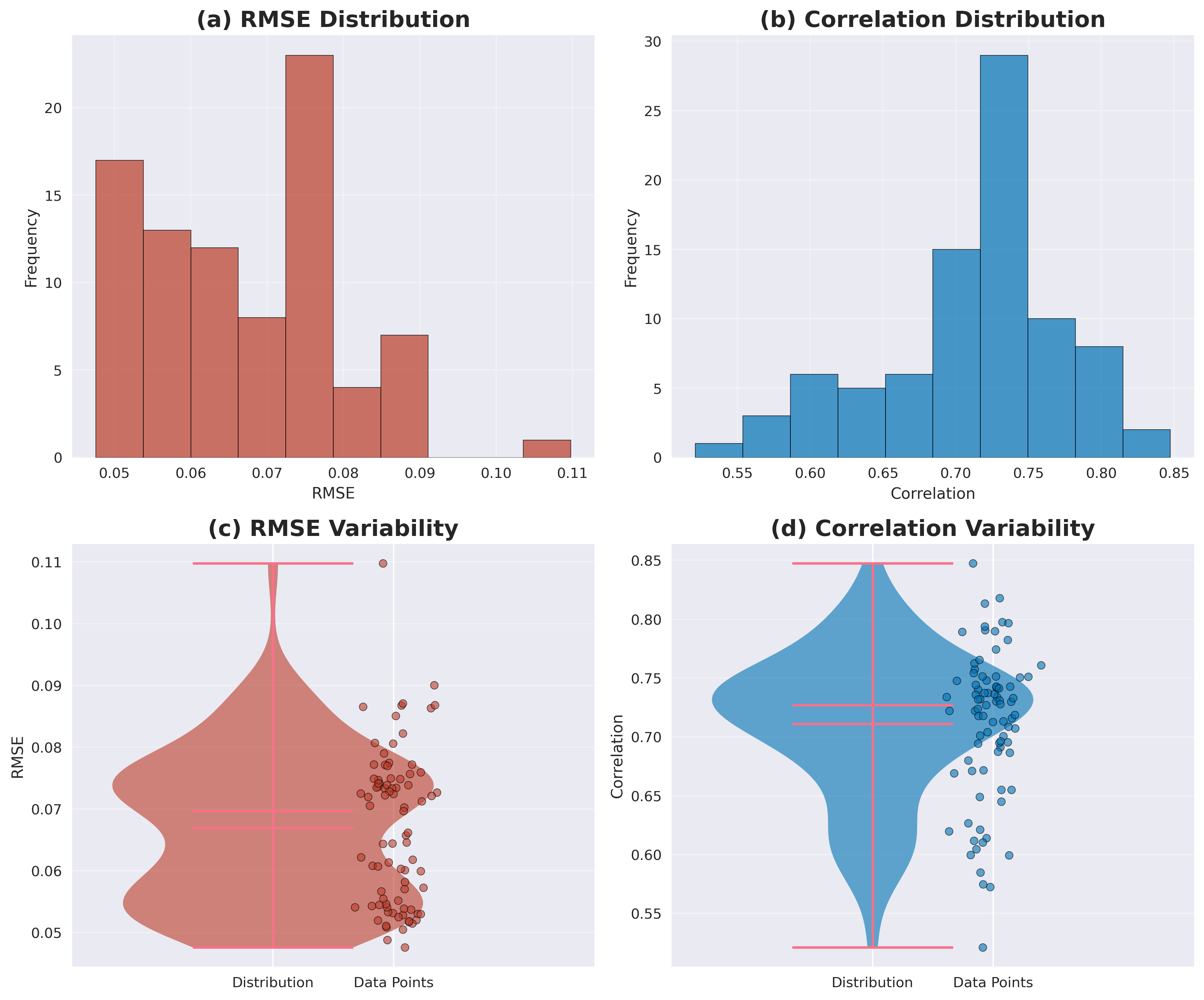}
\caption{(a--b) Histograms and (c--d) violin plots showing national-scale frequency distributions and variability of model performance metrics (RMSE and correlation).}
\label{fig:metrics-distribution}
\end{figure}

For the national-scale validation, the spatial distribution of SWEB's performance metrics across the Australia (Figure~\ref{fig:metrics-map}) shows distinct national-scale patterns related to climate zones and landscape characteristics. RMSE values (panel a) display a clear longitudinal gradient with lowest errors (0.05--0.07) in western regions and higher errors (0.09--0.11) in southeast Australia. Correlation coefficients (panel b) show good performance (0.70--0.85) across most regions with highest correlations in southeast Australia, suggesting regional differences in model structure adequacy. Model performance metric distributions at the national scale are displayed in Figure~\ref{fig:metrics-distribution}. RMSE distribution (panel a, c) is dual-modal and centre-skewed with most values concentrated between 0.05--0.08, indicating consistently good model accuracy across diverse Australian conditions. Correlation distribution (panel b, d) shows a strong peak around 0.75 and minimal values below 0.55, demonstrating robust relationships across the continent.

\section{Implications for potential yield estimation}

Having access to PAW estimates in near real-time is useful for supporting decision making. Nowcasts of RZSM made at any plant growth stage can be used to predict potential yield using the F\&S water-use-efficiency equation \citep{french_water_1984}. The F\&S equation is a simple and widely used method for predicting potential grain yield by relating grain yield, seasonal rainfall, and crop water use. In the presentation, we will explore the application of soil water nowcasts from SWEB in the F\&S potential yield equation to forecast potential grain yield at different points throughout the growing season. We will also present how these water-limited potential yield estimates could then be used to calculate nutrient replacement budgets and provide nutrient prescription maps. Near-real-time updates to soil water estimates are useful for tailoring management decisions that respond to changing seasonal conditions, whether by applying more fertiliser in good years, or adjusting prescriptions to better-match changing conditions.

\newpage

\section{Conclusion}

In this paper, we present SWEB, a practical and scalable modelling framework for monitoring crop water status across Australia's grain-growing regions by explicitly linking surface energy processes with subsurface soil water dynamics. By integrating high-resolution satellite observations, nationally consistent climate data, and physically based soil representations, SWEB provides spatially explicit estimates of $\mathrm{ET}_{a}$ and RZSM at paddock-relevant scales. A national-scale validation against multiple independent SM monitoring networks shows that SWEB performs robustly across diverse climates, soils, and management systems, supporting confidence in its national applicability. Importantly, near-real-time estimates of PAW generated by SWEB create a direct pathway to yield forecasting, nutrient management, and risk-based decision making in water-limited environments. As such, SWEB aligns with GRDC priorities by enabling improved understanding and management of spatial water constraints on crop productivity, supporting more responsive and efficient farm management under variable seasonal conditions.

\section{Acknowledgements}

The research undertaken as part of this project is made possible by the significant contributions of growers through both trial cooperation and the support of the GRDC---the authors would like to thank them for their continued support. The authors also would like to thank all of the custodians of the soil probe networks for making their data freely available for this research.

\bibliographystyle{elsarticle-harv}
\bibliography{references}

\section{Contact details}

Dr Yi Yu\\
The University of Sydney, NSW 2006\\
0449193587\\
\href{mailto:yi.yu1@sydney.edu.au}{yi.yu1@sydney.edu.au}

Prof. Thomas F. A. Bishop\\
The University of Sydney, NSW 2006\\
0405023457\\
\href{mailto:thomas.bishop@sydney.edu.au}{thomas.bishop@sydney.edu.au}

\section{Date published}

February 2026

\end{document}